\documentclass[secnumarabic,amsmath,amssymb, nobibnotes, aps, prl, preprint]{revtex4-1}
\usepackage{graphicx}
\usepackage{natbib}
\usepackage{xfrac}

\begin{document}

\title{Investigation of the concept of beauty via a lock-in feedback experiment}%

\author{M. Kaptein}%
\affiliation{Statistics and Research Methods, Tilburg University, The Netherlands}
\author{R. van Emden}%
\affiliation{Statistics and Research Methods, Tilburg University, The Netherlands}
\affiliation{VU University Amsterdam, Department of Physics and Astronomy and LaserLab, Amsterdam, The Netherlands}
\author{D. Iannuzzi}%
\email[]{d.iannuzzi@vu.nl}
\affiliation{VU University Amsterdam, Department of Physics and Astronomy and LaserLab, Amsterdam, The Netherlands}
\date{\today}%
\begin{abstract}
Lock-in feedback circuits are routinely used in physics laboratories all around the world to extract small signals out of a noisy environment. In a recent paper (M. Kaptein, R. van Emden, and D. Iannuzzi, \emph{paper under review}), we have shown that one can adapt the algorithm exploited in those circuits to gain insight in behavioral economics. In this paper, we extend this concept to a very subjective socio-philosophical concept: the concept of beauty. We run an experiment on 7414 volunteers, asking them to express their opinion on the physical features of an avatar. Each participant was prompted with an image whose features were adjusted sequentially via a lock-in feedback algorithm driven by the opinion expressed by the previous participants. Our results show that the method allows one to identify the most attractive features of the avatar.
\end{abstract}
\maketitle

Let us imagine the following situation. The members of a group of non-interacting individuals are asked to take a decision that involves a certain degree of subjectivity. Their preference can be partially influenced by a set of parameters that are controlled by an external agent who wants to maximize the likelihood of obtaining a certain response. As soon as the external agent tries to vary the parameters to achieve her goal, she realizes that the correlation between causes and effects is hidden in the noise generated by the complexity of the individual decision making process. Without a clear understanding of the effect of her action on the preferences of the respondents, she will probably end up choosing sub-optimal conditions. The problem can be addressed via multiple methods already reported in the literature (see, for example, \cite{NIPS2011_4475} and references therein), a review of which is out of the aim of this paper. It is however interesting to note that, from a logical perspective, the situation is not much different than that routinely experienced in many physics laboratories all around the world, where scientists are confronted with the necessity to extract information on the correlation between physical variables that are immersed in a high level of noise. In physics, this problem is often solved by switching to the frequency domain, where the use of lock-in amplifiers allows one to distinguish even extremely small signals (see, for example, \cite{devore2016improving} and references therein). Tantalized by this analogy, we have recently investigated whether lock-in algorithms may find applications in behavioral economics~\cite{Kaptein2016decoy}. Our experiment clearly demonstrated that, working in the frequency domain, one can indeed gain a more complete understanding of the cause-effect relation in a decision making process. In this paper, we further elaborate on that approach and ask whether lock-in algorithms can be applied to socio-philosophical concepts. Focusing on the concept of beauty, we demonstrate that one can use a lock-in feedback algorithm to quantitatively determine the most attractive facial features of an avatar.

\vspace{12pt}

\begin{figure}
	\begin{center}
		\includegraphics[width=\columnwidth]{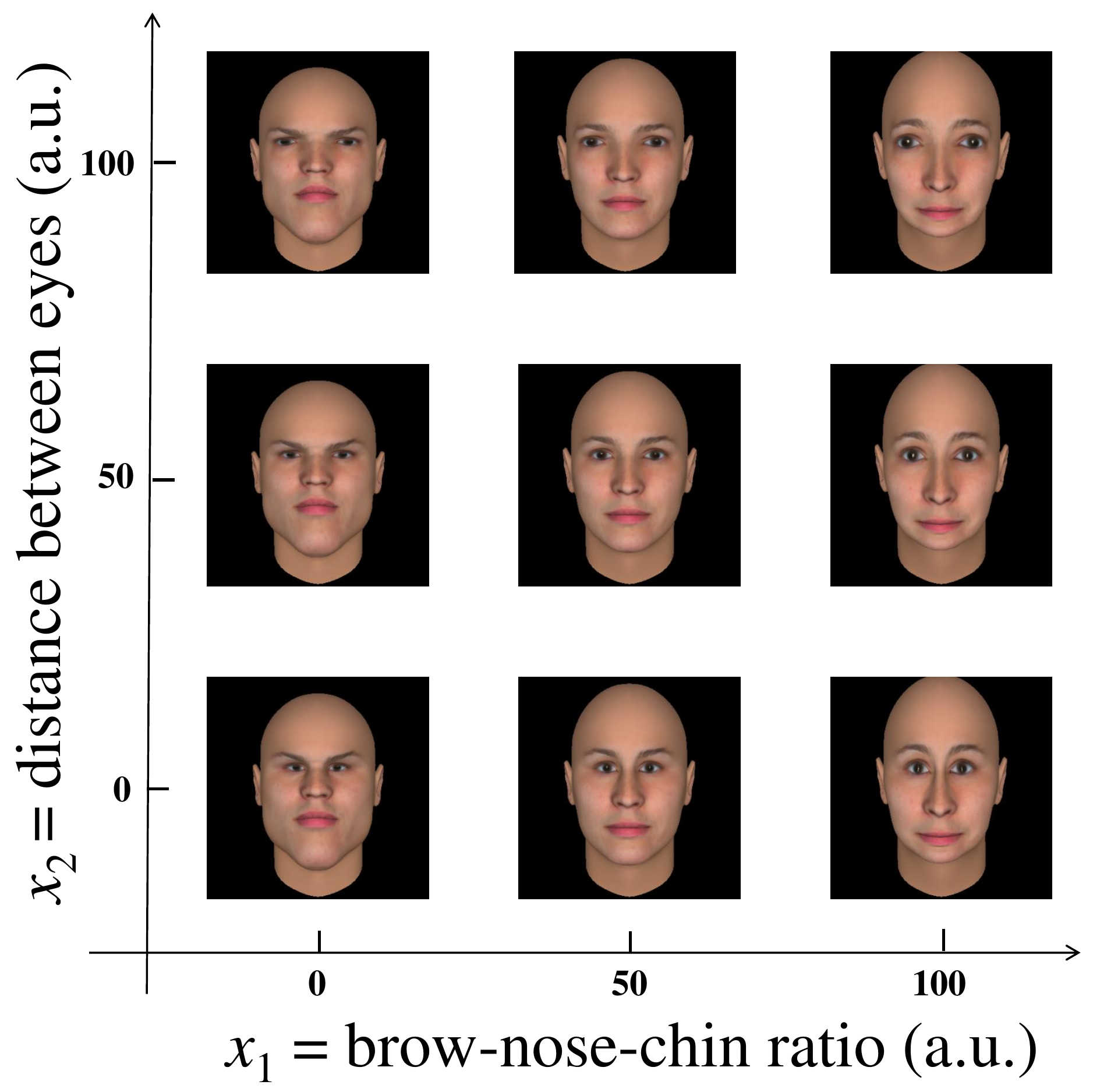}
		\caption{Schematic representation of the attribute space used for this paper. Each of the avatars reported in this graph was obtained from the central one by either increasing or decreasing the elongation of the face ($x_1$) or the distance between the eyes ($x_2$).}
		\label{fig:faces}
	\end{center}
\end{figure}

\noindent \emph{Problem definition:}
Fig.~\ref{fig:faces} shows nine avatars created from an anatomically correct multilayered 3D model of the human face \cite{inversions2003facegen}. The nine figures differ on two parameters: the \emph{brow-nose-chin ratio}, $x_1$, and the \emph{distance between the eyes}, $x_2$. It is know that, in general, the central face is considered as more attractive than the others \cite{galton1879composite}. Our aim is to use a lock-in feedback algorithm to find the values of $x_1$ and $x_2$ that give rise to the most attractive avatar.

\vspace{12pt}

\begin{figure}
	\begin{center}
        \includegraphics[width=\columnwidth]{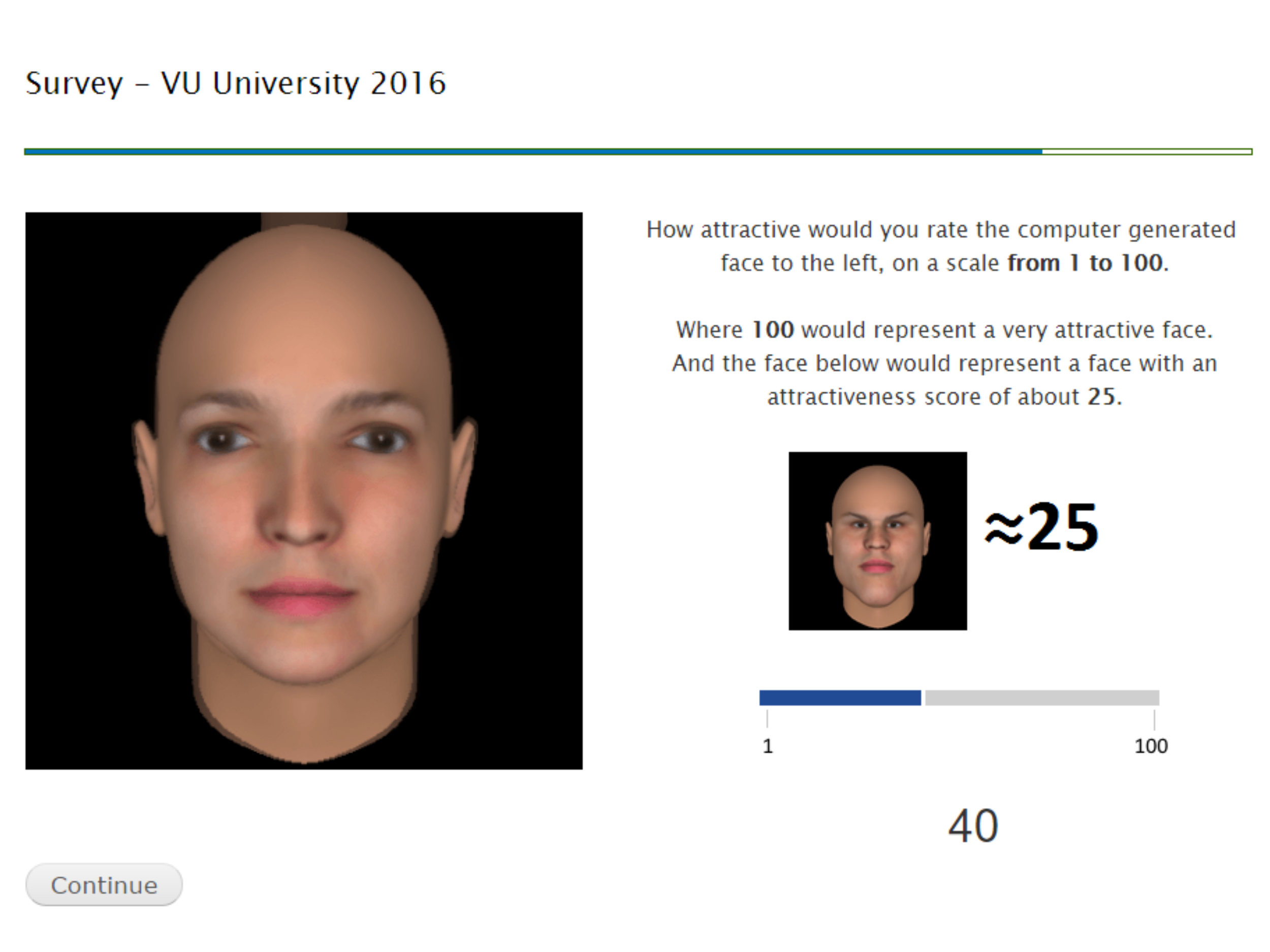}
		\caption{Example of the webpage as prompted to the people who participated to our experiment. All the features of the page remained unaltered throughout the experiment, except for the left avatar, whose brow-nose-chin ratio and eye-to-eye distance were adjusted according to our lock-in feedback algorithm. Participants could express their opinion via the slider on the bottom.}
		\label{fig:web}
	\end{center}
\end{figure}

\noindent \emph{Experimental setup:}
Our experiment was performed on Amazon-Turk -- a web-based tool that has been already recognized as a trustable platform for social science experiments \cite{buhrmester2011amazon,paolacci2010running}. Participants could log-in, perform the task requested, and receive a monetary compensation for their participation. None of them was allowed to log-in more than once. Fig.~\ref{fig:web} shows a print-screen of our experiment. The text and the right avatar remained unaltered throughout the entire duration of the experiment. The elongation of the face and the distance between the eyes of the left avatar, on the contrary, were adjusted sequentially according to the algorithm discussed below. Participants could use the slider to express their opinion.

\vspace{12pt}

\noindent \emph{Description of the algorithm:}
From the argument reported above, it is licit to assume that there exist a value of $x_1$ and a value of $x_2$ for which the appearance of the avatar will maximize the average score $y$ -- the number that, in average, participants will choose via the slider of Fig.~\ref{fig:web}. We will indicate those two maximizing values with $x_{1M}$ and $x_{2M}$. Our goal is to find those two \emph{a priori} unknown values via a lock-in negative feedback loop.

For the sake of simplicity, we will assume that, close to $x_{1M}$ and $x_{2M}$:
\begin{equation}
y(x_1,x_2)= A_1 \left( x_1 - x_{1M} \right)^2 + y_{10} + A_2 \left( x_2 - x_{2M} \right)^2 + y_{20}
\label{eq:parabola}
\end{equation}
\noindent where $A_{1}$, $A_{2}$, $y_{10}$, and $y_{20}$ are unknown constants. Let us suppose that the values of $x_1$ and $x_2$ as seen by the $i^{th}$ participant are tuned according to:
\begin{equation}
x_{1,i}=\widetilde{x}_{1,i} + \delta_1 \cos \left( \omega_1 i \right)
\label{eq:oscillation1}
\end{equation}
\begin{equation}
x_{2,i}=\widetilde{x}_{2,i} + \delta_2 \cos \left( \omega_2 i \right)
\label{eq:oscillation2}
\end{equation}

\noindent where $i$ goes from 1 to the total number of participants $N$; $\widetilde{x}_{1,1}$, $\widetilde{x}_{2,1}$, $\omega_1$, $\omega_2$, $\delta_1$, and $\delta_2$ are six suitably chosen constants set at the start of the experiment; and  $\widetilde{x}_{1,i}$ and $\widetilde{x}_{2,i}$ have to be sequentially adjusted to find the value of $x_{1M}$ and $x_{2M}$. Plugging eq.~\ref{eq:oscillation1} and eq.~\ref{eq:oscillation2} into eq.~\ref{eq:parabola}, one can conclude that the expected response of the $i^{th}$ participant is given by:

\begin{equation}
\begin{aligned}
y^{expected}_i= & A_1 \left( \widetilde{x}_{1,i} + \delta_1 \cos \left( \omega_1 i \right) - x_{1M} \right)^2 + y_{10} + \\
& A_2 \left( \widetilde{x}_{2,i} + \delta_2 \cos \left( \omega_2 i \right) - x_{2M} \right)^2 + y_{20} + \gamma_i\\
\end{aligned}
\label{eq:parabolaplusoscillation}
\end{equation}

\noindent where we have added the term $\gamma_i$ to include the noise generated by the personal preference of the $i^{th}$ participant. Eq.~\ref{eq:parabolaplusoscillation} yields:

\begin{equation}
\begin{aligned}
y_i^{expected}= & 2 A_1 \left(\widetilde{x}_{1,i} - x_{1M} \right) \delta_1 \cos \left( \omega_1 i \right) + \\
& 2 A_2 \left(\widetilde{x}_{2,i} - x_{2M} \right) \delta_2 \cos \left( \omega_2 i \right) + \\
& A_1 \left( \widetilde{x}_{1,i} - x_{1M} \right)^2 + A_2 \left( \widetilde{x}_{2,i} - x_{2M} \right)^2 + \\
& \frac{A_1 \delta_1^2 \cos\left( 2\omega_1 i \right)}{2} + \frac{A_2 \delta_2^2 \cos\left( 2\omega_2 i \right)}{2} + \\
 & \frac{A_1 \delta_1^2}{2} + \frac{A_2 \delta_2^2}{2} + y_{10} + y_{20} + \gamma_i
\end{aligned}
\label{eq:simplified}
\end{equation}

\noindent Note that the amplitude of the oscillations at $\omega_1$ is proportional to how far the attribute $x_1$ is from the ideal value. Similarly, the amplitude of the oscillations at $\omega_2$ is proportional to how far the attribute $x_2$ is from the ideal value. One can thus use a lock-in algorithm to isolate these contributions from the others and drive a negative feedback circuit to sequentially bring $\widetilde{x}_1$ and $\widetilde{x}_2$ closer and closer to $x_{1M}$ and $x_{2M}$, respectively.

Following this approach, at the start of the experiment we first collect the value of $y$ for the first $n_1$ participants, where $n_1$ is a constant number set \emph{a priori}, with $n_1<<N$. During this first phase, $\widetilde{x}_{1,i}$ is kept constant: $\widetilde{x}_{1,1...n_1}=\widetilde{x}_{1,1}$. For each value of $i$ from $1$ to $n_1$, we multiply the experimental value of $y$ times $\cos \left( \omega_1 i \right)$, and sum the resulting products from $i=1$ to $i=n_1$:

\begin{equation}
y^{exper}_{lock1,n_1} = \sum_{i=1}^{n_1} y^{exper}_i \cos \left( \omega_1 i\right)
\label{eq:lock_1}
\end{equation}

\noindent Following the working principle of negative feedback loops, we can now use the result of eq. \ref{eq:lock_1} to set the value of $\widetilde{x}_{n_1+1}$:

\begin{equation}
\widetilde{x}_{1,n_1+1}= \frac{\sum_{i=1}^{n_1}\widetilde{x}_{1,i}}{n_1} - \gamma_1 y^{exper}_{lock1,n_1}
\label{next_value_bis}
\end{equation}

\noindent where $\gamma_1$ is a constant that we fixed \emph{a priori}. Then, after that the $(n_1+1)^{th}$ participant has answered, we calculate the summation of eq.~\ref{eq:lock_1} and eq.~\ref{next_value_bis} for $i$ that goes from $2$ to $n_1+1$, and apply the same procedure to determine the values of $\widetilde{x}_{1,n_1+2}$. Iterating the procedure further via the generic equations:

\begin{equation}
y^{exper}_{lock1,j} = \sum_{i=j-n_1+1}^{j} y^{exper}_i \cos \left( \omega_1 i\right)
\label{eq:lock_1_gen}
\end{equation}

\begin{equation}
\widetilde{x}_{1,j+1}= \frac{\sum_{i=j-n_1+1}^{j}\widetilde{x}_{1,i}}{n_1} - \gamma_1 y^{exper}_{lock1,j}
\label{next_value_bis_gen}
\end{equation}

\noindent one should observe that the value of $\widetilde{x}_{1,i}$ eventually reaches $x_{1M}$, where it should remain locked until the end of the experiment. Applying, in parallel, a similar algorithm to the variable $x_2$, one can simultaneous bring $\widetilde{x}_{2,i}$ to $x_{2M}$.

To understand why the negative feedback loop described above should converge to the optimal values, one can calculate the expected signal that the lock-in algorithm should give if the experimental values of $y$ followed exactly the expected trend ($y_i^{exper}=y_i^{expected}$). Plugging eq.~\ref{eq:simplified} into eq.~\ref{eq:lock_1_gen}, one obtains:

\begin{equation}
y^{expected}_{lock1,j} = A_1 \delta_1 \sum_{i=j-n_1+1}^{n_1} \left( \widetilde{x}_{1,i} - x_{1M} \right) + o.t.
\label{lock_2}
\end{equation}

\noindent where $o.t.$ indicates terms that, for a sufficiently large value of $n_1$, become negligible. Inverting eq.~\ref{lock_2}, one can indeed verify that:

\begin{equation}
x_{1M} \approx \frac{\sum_{i=j-n_1+1}^{n_1} \widetilde{x}_{1,i}}{n_1} - \frac{y^{expected}_{lock1,j}}{A_1 \delta_1 n_1}
\label{lock_2_final}
\end{equation}

\noindent For a suitable choice of $\gamma_1$, $\gamma_2$, $\delta_1$, and $\delta_2$, the algorithm presented should thus be able to complete the task.

\vspace{12pt}

\begin{figure}
	\begin{center}
		\includegraphics[width=\columnwidth]{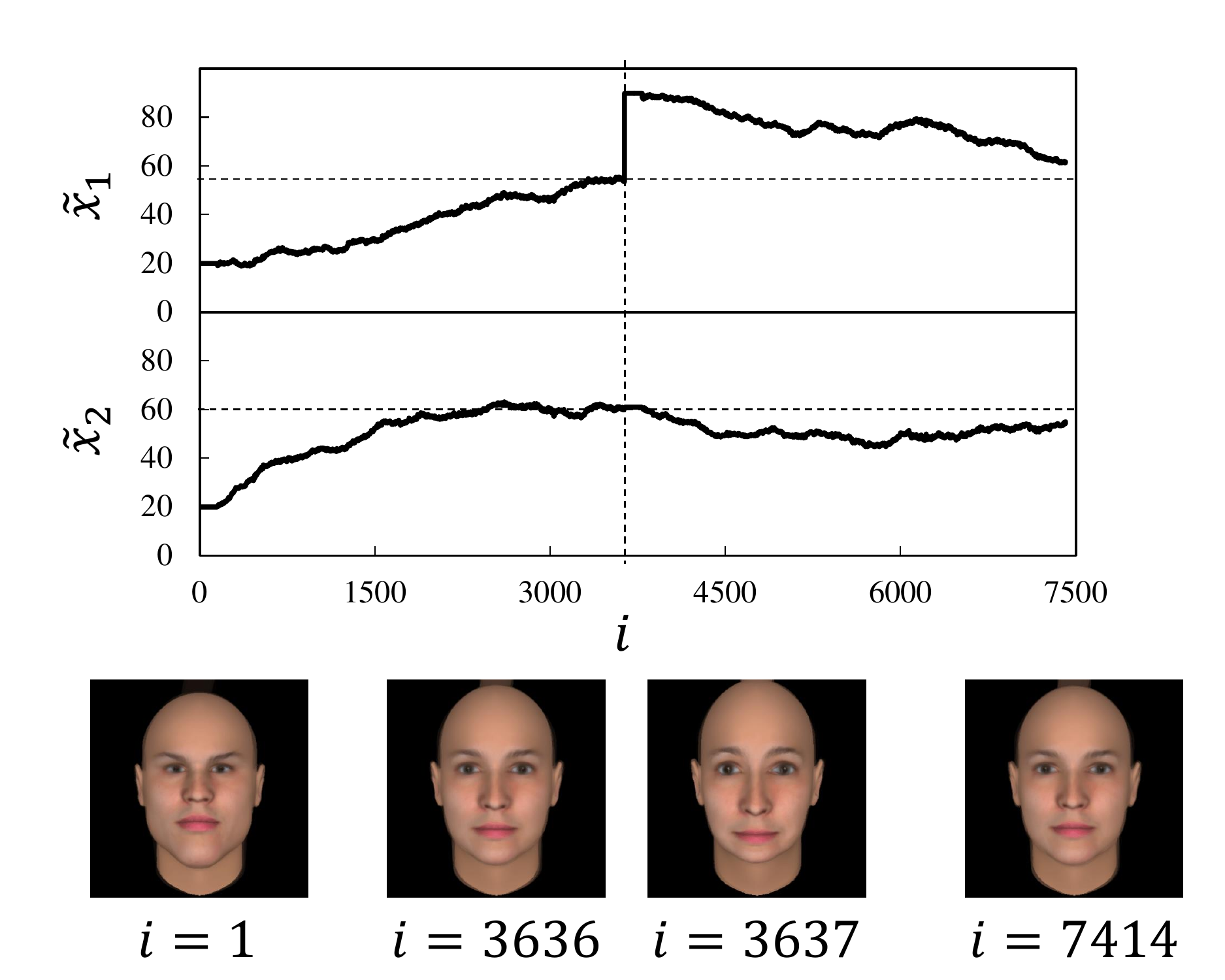}
		\caption{Evolution of $\tilde{x}_1$ and $\tilde{x}_2$ as a function of the participant number $i$. The vertical dashed lines indicates the instant in which we forced $\tilde{x}_1=90$ ($i=3637$). The two horizontal lines indicate the values of $\tilde{x}_1$ and $\tilde{x}_2$ that optimize the avatar's appearance as obtained from the first phase of the experiment. The avatars below the graph show the starting and arriving points of the two phases of the experiment.}
		\label{fig:res}
	\end{center}
\end{figure}

\noindent \emph{Experimental results:}
Fig.~\ref{fig:res} shows the results of our experiment, which was performed using the parameters reported in table~\ref{tab:parameters}. The values of $x_1$ and $x_2$ were parameterized on a linear scale from 0 to 100 according to Fig.~\ref{fig:faces}. In the first phase of the experiment, we set $\tilde{x}_{1,1}=20$ and $\tilde{x}_{2,1}=20$, and let the program run until $i=3636$. We then moved $\tilde{x}_{1,3637}$ to $90$, and observed the lock-in feedback recovering from this perturbation until $i=N=7414$.

\begin{table}
\caption {Initial parameters.} \label{tab:parameters}
\begin{tabular}{|l|c|}
  \hline
  Lock-in $1$ & $\omega_1=2.63$; $n_1=150$; $\delta_1=8$; $\gamma_1=0.0006$ \\  \hline
  Lock-in $2$ & $\omega_2=2.51$; $n_2=150$; $\delta_2=8$; $\gamma_2=0.0006$ \\  \hline
\end{tabular}
\end{table}

Clearly, in the first part of the experiment ($i$ from $1$ to $3636$), the algorithm leads to $\tilde{x}_1 \approx 55$ and $\tilde{x}_2 \approx 60$. These values are in agreement with what has been suggested in the previous literature \cite{langlois1990attractive}. As for the second part of the experiment, it is first interesting to note that, as expected, after the forced change, the variable $\tilde{x}_1$ moves again towards a value close to the one that the previous part of the experiment indicated as optimal. It is also quite remarkable to see that, as soon as we set $\tilde{x}_1=90$, the variable $\tilde{x}_2$, which was already optimized in the first phase, starts to decrease before moving back towards the optimal value. We believe that this behavior is due to the fact that the function that connects $y$ with $x_1$ and $x_2$, which we simplified as the sum of two parabolas in eq.~\ref{eq:parabola}, also involves cross terms that mix the two variables. Hence, the optimal value of $x_2$ actually depends on the current value of $x_1$.

\begin{figure}
	\begin{center}
		\includegraphics[width=\columnwidth]{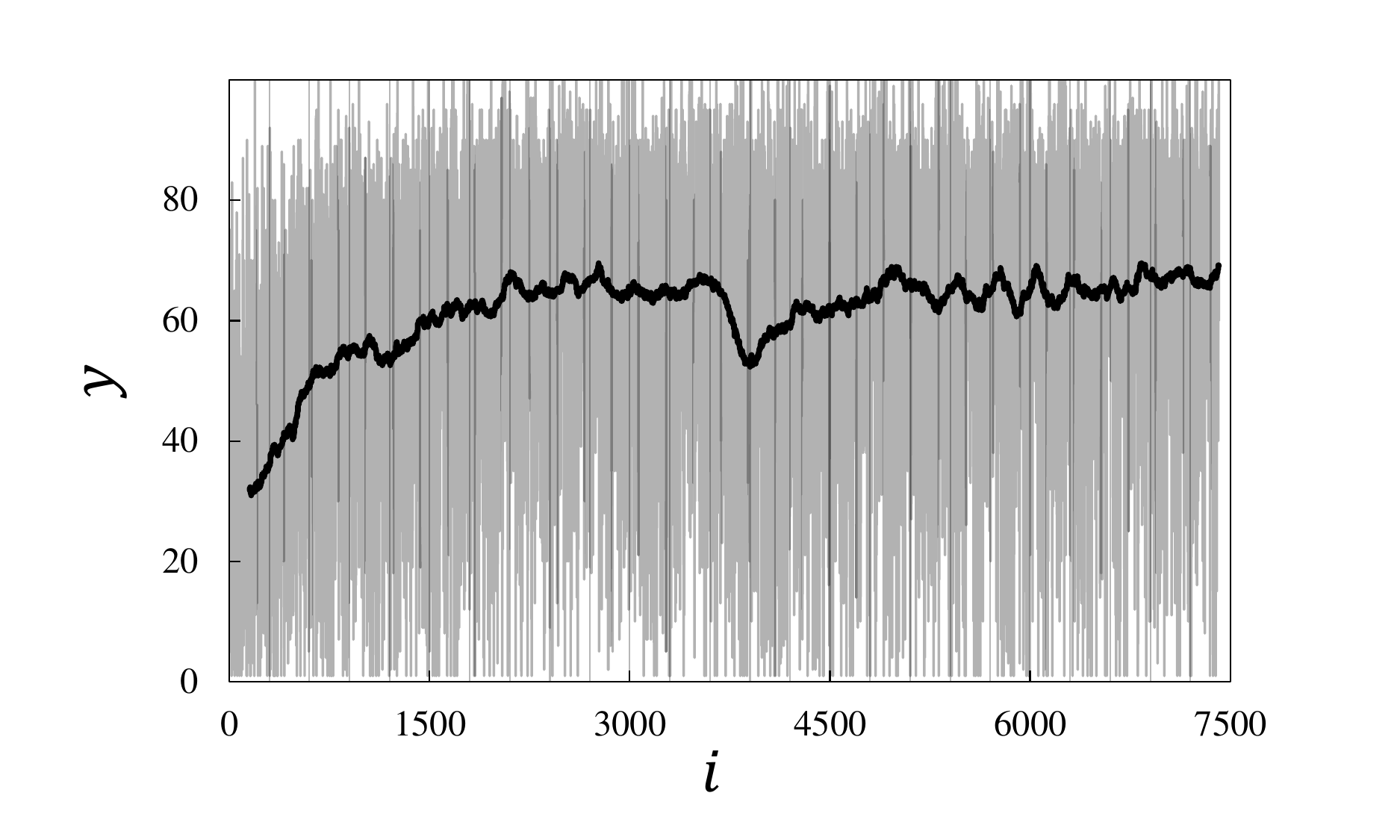}
		\caption{Grey line: Evolution of $y$ as a function of the participant number $i$. Black line: Same data after taking a running average over 150 participants.}
		\label{fig:noise}
	\end{center}
\end{figure}

For the sake of completeness, in Fig.~\ref{fig:noise} we report the evolution of the $y$ as a function of $i$ (grey line), along with a running average performed over a sample of 150 participants. It is evident that the string of raw data only provides very limited information on the dynamics of the experiment. The running average, on the contrary, demonstrates that both phases converge to the same average value of $y$.

\vspace{12pt}

\noindent \emph{Conclusions:}
We have shown that one can adapt the idea of lock-in feedback circuits to determine the physiognomical features that an avatar must have to optimize its aspect. Our experiment paves the way for further applications of lock-in feedback circuits in the social sciences, economics, philosophy, and art.

\vspace{12pt}

\noindent \emph{Acknowledgements:}

The experimental procedure was approved by the Research Ethics Review Board of the Faculty of Economics and Business Administration of the VU Universiteit Amsterdam. DI acknowledges the support of the European Research Council (grant agreement n. 615170) and of the Stichting woor Fundamenteel Onderzoek der Materie (FOM), which is financially supported by the Netherlands Organization for Scientific Research (NWO). We further acknowledge Andrea Giansanti for useful discussions.

\bibliography{z_iannuzzi_ref}

\end{document}